\documentclass[noshowpacs,amsmath,
twocolumn,
superscriptaddress,
8pt,
aps,prb]{revtex4-1}
\usepackage{setspace}
\usepackage{amsmath}
\usepackage{graphicx}
\usepackage[nearskip,margin = 0pt]{subfig}

\usepackage{verbatim}
\usepackage{amsfonts}
\usepackage{amssymb}
\usepackage{epstopdf} 
\usepackage{xcolor}
\usepackage{verbatim}
\usepackage{upgreek}
\usepackage{ragged2e}
\usepackage{url}
\usepackage[hidelinks]{hyperref}
\DeclareGraphicsExtensions{.pdf,.eps,.png,.jpg,.mps}

\begin{document}
\title{Kerr optical frequency division with integrated photonics for stable microwave and mmWave generation}

\author{Shuman Sun$^{1}$, Mark W. Harrington$^{2}$, Fatemehsadat Tabatabaei$^{1}$, Samin Hanifi$^{1}$, Kaikai Liu$^{2}$, Jiawei Wang$^{2}$, Beichen Wang$^{1}$,  Zijiao Yang$^{1,3}$, Ruxuan Liu$^{1}$, Jesse S. Morgan$^{1}$, Steven M. Bowers$^{1}$, Paul A. Morton$^{4}$, Karl D. Nelson$^{5}$, Andreas Beling$^{1}$, Daniel J. Blumenthal$^{2,\dagger}$ and Xu Yi$^{1,3,\dagger}$\\
\vspace{3pt}
$^1$Department of Electrical and Computer Engineering, University of Virginia, Charlottesville, Virginia 22904, USA.\\
$^2$Department of Electrical and Computer Engineering, University of California Santa Barbara, Santa Barbara, California 93016, USA.\\
$^3$Department of Physics, University of Virginia, Charlottesville, Virginia 22904, USA.\\
$^4$ Morton Photonics, Palm Bay, Florida 32905, USA.\\
$^5$ Honeywell International, Plymouth, Minnesota 55441, USA.\\
$^{\dagger}$Corresponding authors: danb@ucsb.edu, yi@virginia.edu}

\begin{abstract}
Optical frequency division (OFD) has revolutionized microwave and mmWave generation and set spectral purity records owing to its unique capability to transfer high fractional stability from optical to electronic frequencies.  Recently, rapid developments in integrated optical reference cavities and microresonator-based optical frequency combs (microcombs) have created a path to transform OFD technology to chip scale. Here, we demonstrate an ultra-low phase noise mmWave oscillator by leveraging integrated photonic components and Kerr optical frequency division. The oscillator derives its stability from an integrated CMOS-compatible SiN coil cavity, and the optical frequency division is achieved spontaneously through Kerr interaction between the injected reference lasers and soliton microcombs in the integrated SiN microresonator.  Besides achieving record-low phase noise for integrated mmWave oscillators, our demonstration greatly simplifies the implementation of integrated OFD oscillators and could be useful in applications of Radar, spectroscopy, and astronomy.
\end{abstract}
\date{\today}

\maketitle

\noindent {\bf Introduction}

Stable microwaves and mmWaves are of critical importance to a wide range of applications, including radar, astronomy, and spectroscopy. Thanks to the invention of optical frequency division (OFD) \cite{fortier2011generation}, the spectral purity of optically generated microwaves and mmWaves have surpassed all other approaches\cite{fortier2011generation,xie2017photonic,nakamura2020coherent,li2023small}. In the OFD, the high fractional stability of optical reference cavities can be coherently transferred from optical to microwave frequencies by using optical frequency combs. When phase locking the frequencies of comb lines to that of the optical reference by feedback control of the comb repetition rate, the frequency and phase of the optical reference are divided down coherently to the comb repetition rate. Stable microwaves can be produced by detecting the comb on a fast photodiode. 
Identical to all frequency divisions, the phase noise of the output is reduced by the square of the division ratio relative to the input signal. Recently, the rapid developments in integrated optical references \cite{lee2013spiral,jin2021hertz,li2021reaching,liu202236} and soliton microresonator-based frequency combs (microcombs) \cite{herr2014temporal,brasch2016photonic,kippenberg2018dissipative} provide a path to miniaturize optical frequency division to the chip scale \cite{tetsumoto2021optically,sun2023integrated,kudelin2023photonic,zhao2023all}. The low size, weight, and power (SWaP) of integrated OFD oscillators will extend the OFD technology to applications in the fields that demand exceedingly low microwave phase noise and portability. Furthermore, the repetition rates in soliton microcombs can reach mmWave and THz frequencies \cite{zhang2019terahertz,tetsumoto2021optically,wang2021towards,sun2023integrated}, which are important to applications in 5G/6G wireless communications \cite{rappaport2019wireless}, radio astronomy \cite{clivati2017vlbi}, and radar \cite{ghelfi2014fully}.

\begin{figure*}[!bht]
\captionsetup{singlelinecheck=off, justification = RaggedRight}
\includegraphics[width=17cm]{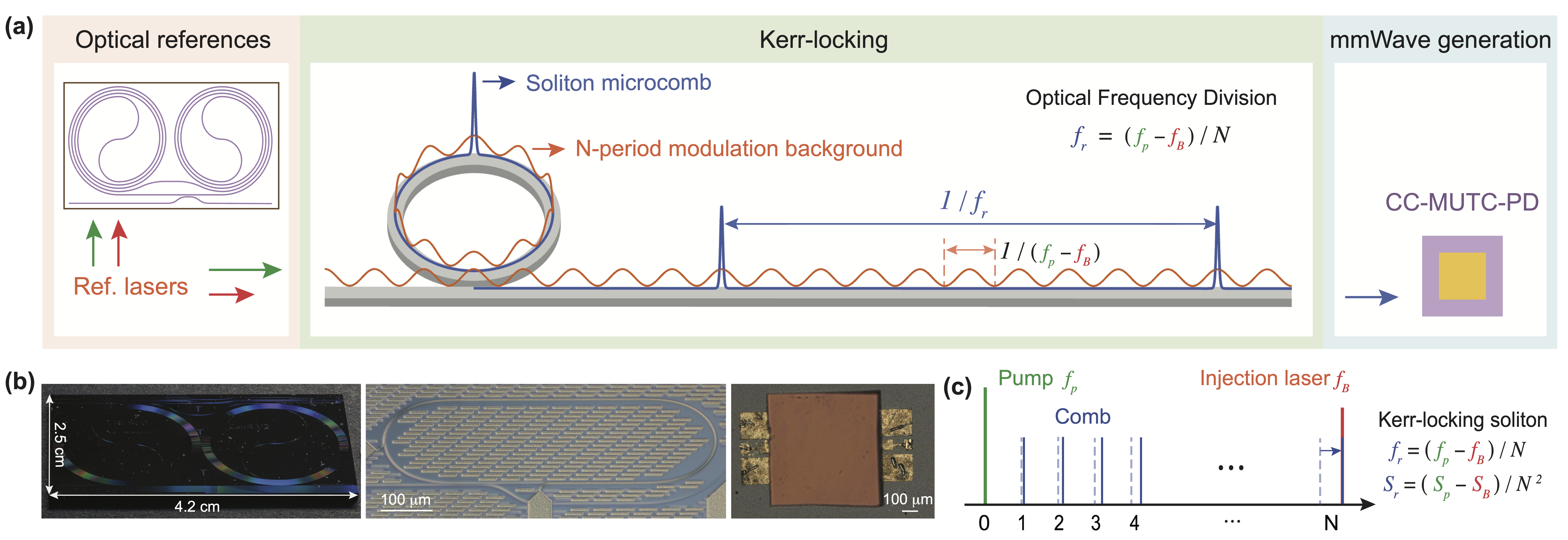}
\caption{{\bf Concept of Kerr optical frequency division for stable microwave and mmWave generation.} {\bf (a)} Conceptual schematic of Kerr OFD. A pair of reference lasers (A and B) are created by stabilizing them to an integrated SiN coil reference cavity. Reference laser A is used to pump an integrated SiN microresonator to generate soliton microcomb, and it also serves as the 0-th comb line. Reference laser B is injected into the microresonator, and its frequency is near the frequency of the $N$-th comb line. Inside the microresonator, reference lasers A and B create amplitude modulation background, which traps soliton through the Kerr effect and synchronizes the timing of the soliton microcomb with the reference laser pair. In the time domain, the soliton repetition period is $N$ times of the reference laser beat note period. In the frequency domain, the soliton repetition rate is $1/N$ of the frequency difference of the two reference lasers. As a result, the reference laser frequency difference is divided down by $N$ to the soliton repetition rate. Photodetecting the soliton on a fast CC-MUTC PD generates stable microwave and mmWaves. {\bf (b)} Pictures of key photonic elements in the Kerr OFD. Left: an integrated 4-meter SiN coil reference cavity. Middle: an integrated SiN microresonator with 109.5 GHz FSR. Right: a typical flip-chip bonded CC-MUTC PD. {\bf (c)} Frequency domain illustration of Kerr OFD. When the soliton is Kerr locked to the reference lasers, its $N$-th comb line moves from its free-running frequency (dashed line) to the injection laser frequency through the change of comb repetition rate. The $S$ is the phase noise.}
\label{fig:concept}
\end{figure*}

Here, we report a record low phase noise mmWave generation based on integrated photonics \cite{sun2023integrated} and Kerr-induced optical frequency division (Kerr OFD) \cite{wildi2023sideband,moille2023kerr,zhao2023all}. In contrast to traditional OFD, there is no servo-control-loop in the Kerr OFD. Instead, the stable optical reference lasers are circulating in the soliton microcomb resonator to directly synchronize the timing of the soliton through Kerr effect. While Kerr optical frequency division has been recently proposed\cite{wildi2023sideband,moille2023kerr,zhao2023all}, its combination with ultra-stable optical references has not been shown yet. In our oscillator, the phase stability is provided by a pair of stabilized lasers at 1551 nm and 1600 nm that are locked to a common CMOS-compatible, integrated SiN coil reference cavity \cite{liu202236,sun2023integrated}. The reference laser at 1551 nm ($f_A$) is amplified to pump the microresonator to generate soliton microcombs, and it serves as the 0-th comb line. The other reference laser at 1600 nm ($f_B$) is injected into the microresonator, and its frequency is tuned to near the $N$-th comb line. The two reference lasers create periodic intensity modulation inside the microresonator, which transforms into periodic refractive index modulation through the Kerr effect and traps soliton to the highest refractive index locations \cite{jang2015temporal,lu2021synthesized}. This synchronizes the soliton timing with the reference lasers in the time domain, while in the frequency domain, 
the frequency difference of the two reference lasers is divided by $N$ times to the soliton repetition rate, $f_r$. The soliton is then detected on a high-speed flip-chip bonded charge-compensated modified uni-traveling carrier photodiode (CC-MUTC PD) \cite{xie2014improved,wang2021towards} to generate 109.5 GHz mmWaves with exceedingly low phase noise. At 100 Hz and 10 kHz offset frequency, the phase noise reaches $-77$ dBc/Hz and -121 dBc/Hz, respectively, which correspond to $-98$ dBc/Hz and $-142$ dBc/Hz when carrier frequency is scaled to 10 GHz. To the best of our knowledge, this is the lowest phase noise for an integrated photonic mmWave oscillator.  

\begin{figure*}[!bht]
\captionsetup{singlelinecheck=off, justification = RaggedRight}
\includegraphics[width=17cm]{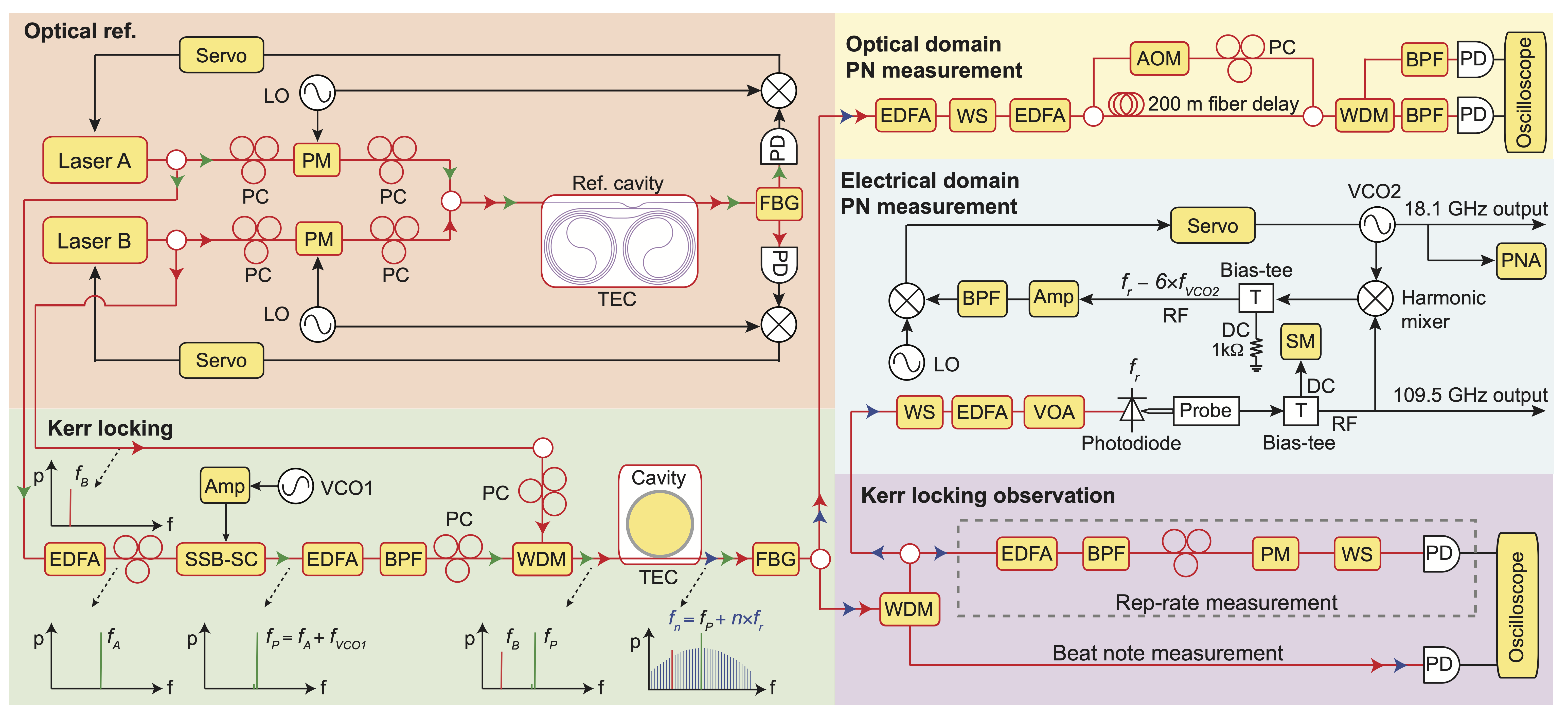}
\caption{{\bf Experimental setup.} Laser A and laser B are stabilized to the coil reference cavity and serve as reference lasers. Laser A is used to create the pump laser that generates solitons in an integrated microresonator. Laser B is combined with the pump laser on a WDM and injected into the same microresonator for Kerr locking and Kerr OFD. Both the coil reference cavity and the soliton microresonator are temperature-controlled by thermoelectric coolers (TEC). The soliton microcomb is then split for (i) optical domain phase noise (PN) measurement, (ii) mmWave generation and electrical domain phase noise measurement and (iii) Kerr locking phenomenon observation. Erbium-doped fiber amplifiers (EDFAs), polarization controllers (PCs), phase modulators (PMs), single-sideband modulator (SSB-SC), band pass filters (BPFs), fiber Bragg grating (FBG) filters, line-by-line waveshaper (WS), acoustic-optics modulator (AOM), phase noise analyzer (PNA), electrical amplifiers (Amps), source meter (SM) and variable optical attenuator (VOA) are used in the experiment.}
\label{fig:setup}
\end{figure*}


It is important to compare this Kerr OFD approach with the conventional servo-controlled OFD approach. In conventional servo OFDs, the phase error between the comb line and the optical reference is detected and is corrected by feedback control of a current or voltage that tunes the comb repetition rate. Auxiliary optoelectronic components, such as photodiodes, optical amplifiers, low-frequency local oscillator, electronic PID controls, are often necessary for the conventional OFD \cite{fortier2011generation,tetsumoto2021optically,sun2023integrated,kudelin2023photonic}. In addition, OFD requires large servo bandwidth, and thus the bandwidth of all auxiliary components and the comb repetition rate tuning mechanism have to be high. 
To date, the servo bandwidths of microcomb-based OFD are limited to hundreds of kHz \cite{tetsumoto2021optically,kudelin2023photonic,sun2023integrated,jin2024microresonator}. In contrast, the Kerr OFD leverages the strong Kerr interaction between the reference lasers and temporal solitons in the optical microresonator, and the soliton is passively locked to the optical references. No auxiliary component is needed, which greatly simplifies the implementation of optical frequency division. In addition, large locking bandwidth is feasible with small injection power due to the cavity-enhanced Kerr interaction. In our demonstration, when injecting only 185 $\mu$W optical power of the reference laser into the cavity, the maximum locking bandwidth is estimated to reach 30s MHz. 
We believe the demonstration of Kerr OFD paves the path for high-performance, fully-integrated low noise photonic microwave and mmWave oscillators.

\medskip

\noindent {\bf Results.} 

\begin{figure*}[!bht]
\captionsetup{singlelinecheck=off, justification = RaggedRight}
\includegraphics[width=17cm]{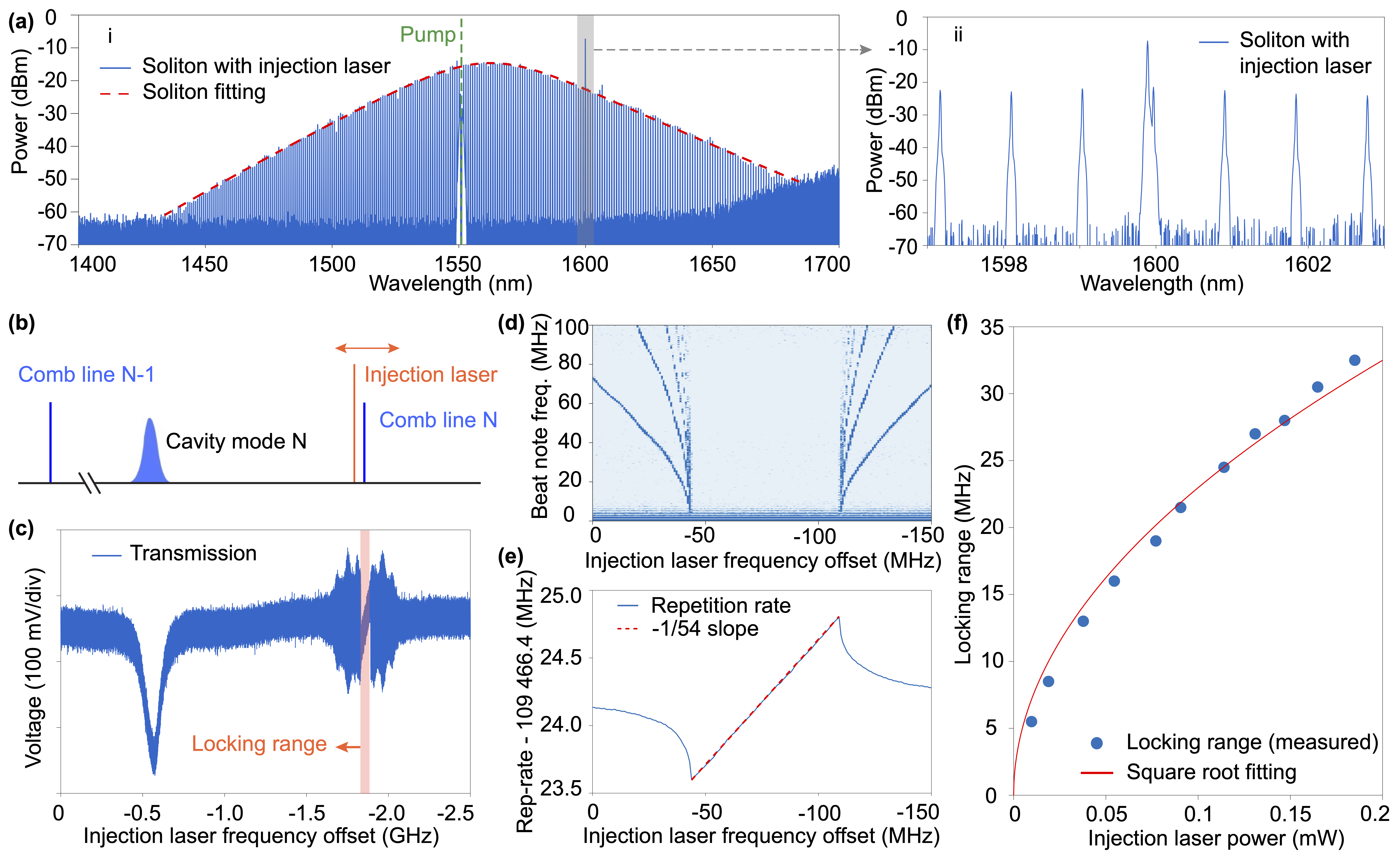}
\caption{{\bf Observation of Kerr locking and Kerr optical frequency division.} {\bf (a)} Optical spectrum of soliton microcombs and the injection laser. A zoom-in figure of the spectrum around the injection laser wavelength is shown on the right. {\bf (b)} Illustration of the measurement in the frequency domain. The frequency of the injection laser ($f_B$) is scanned continuously near the $N$-th comb line and $N$-th cavity mode. {\bf (c)} The transmission of the injection laser and the soliton microcomb versus the frequency of the injection laser. When the frequency of the injection laser enters the Kerr locking range (area marked by orange), comb line $N$ is locked to the injection laser and the beat note oscillator disappears and is reduced to a DC voltage.  {\bf (d)} The spectrogram of the locking regime, which gives the instantaneous frequency of the transmission. The beat note frequency disappears in the middle, showing the frequency of comb line $N$ is locked to the injection laser. {\bf (e)} Measured soliton repetition rate versus the injection laser frequency. A red dashed line with a slope of -1/54 is plotted in the locking region, validating the optical frequency division. {\bf (f)} One-sided locking range is measured versus the injection laser power. The measurements match nicely with the squared-root fitting (red dashed).}
\label{fig:Kerr_Lock}
\end{figure*}

\begin{figure*}[!bht]
\captionsetup{singlelinecheck=off, justification = RaggedRight}
\includegraphics[width=17cm]{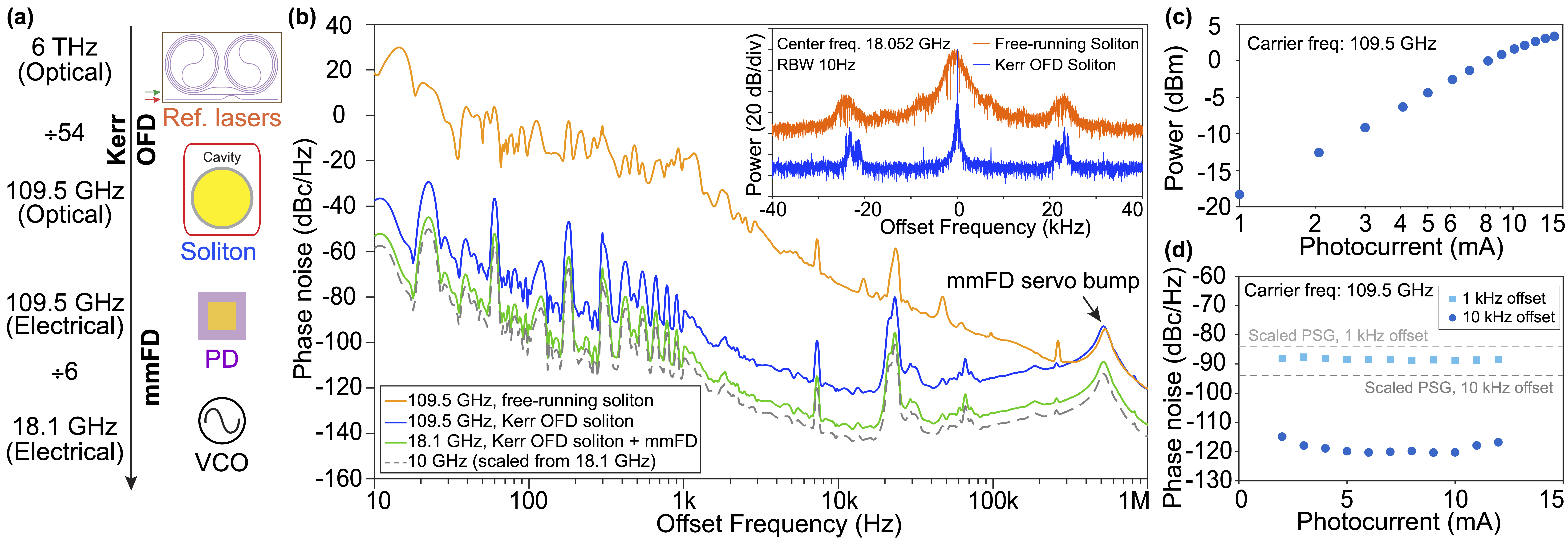}
\caption{{\bf Characterization of mmWave generated from Kerr optical frequency division.} {\bf 
(a)} Illustration of phase noise measurement for the mmWave generated with Kerr-OFD. The 6 THz frequency difference of the two optical reference lasers is divided down to 109.5 GHz through Kerr-OFD, and mmWave is generated on a CC-MUTC PD. To measure the phase noise of the mmWave, a mmWave to microwave frequency division (mmFD) is used to divide the 109.5 GHz output by 6 to the 18.1 GHz output of a VCO, whose phase noise can be directly measured on a phase noise analyzer. {\bf (b)} Phase noise characterization. The phase noise of the mmWave output after mmFD at 18.1 GHz carrier frequency is shown in green. Scaling the carrier frequency back to 109.5 GHz mmWave gives the upper bound of the mmWave phase noise (blue for Kerr OFD soliton, orange for free running soliton). For general comparison, the phase noise is also scaled to a typical 10 GHz (gray-dashed). Inset: electrical spectra of the Kerr OFD output after mmFD for free-running soliton (orange) and Kerr-OFD soliton (blue).  {\bf (c)} The mmWave power versus photocurrents. The maximum output power is 3.4 dBm with photocurrents of 14 mA. {\bf (d)} Phase noise of mmWave at 1 kHz and 10 kHz offset versus photocurrents. For comparison, the phase noise of the Keysight PSG standard model with its carrier frequency scaled to 109.5 GHz is shown in dashed gray lines.}
\label{fig:phase_noise}
\end{figure*}


\begin{figure}[!bht]
\captionsetup{singlelinecheck=off, justification = RaggedRight}
\includegraphics[width=8.5cm]{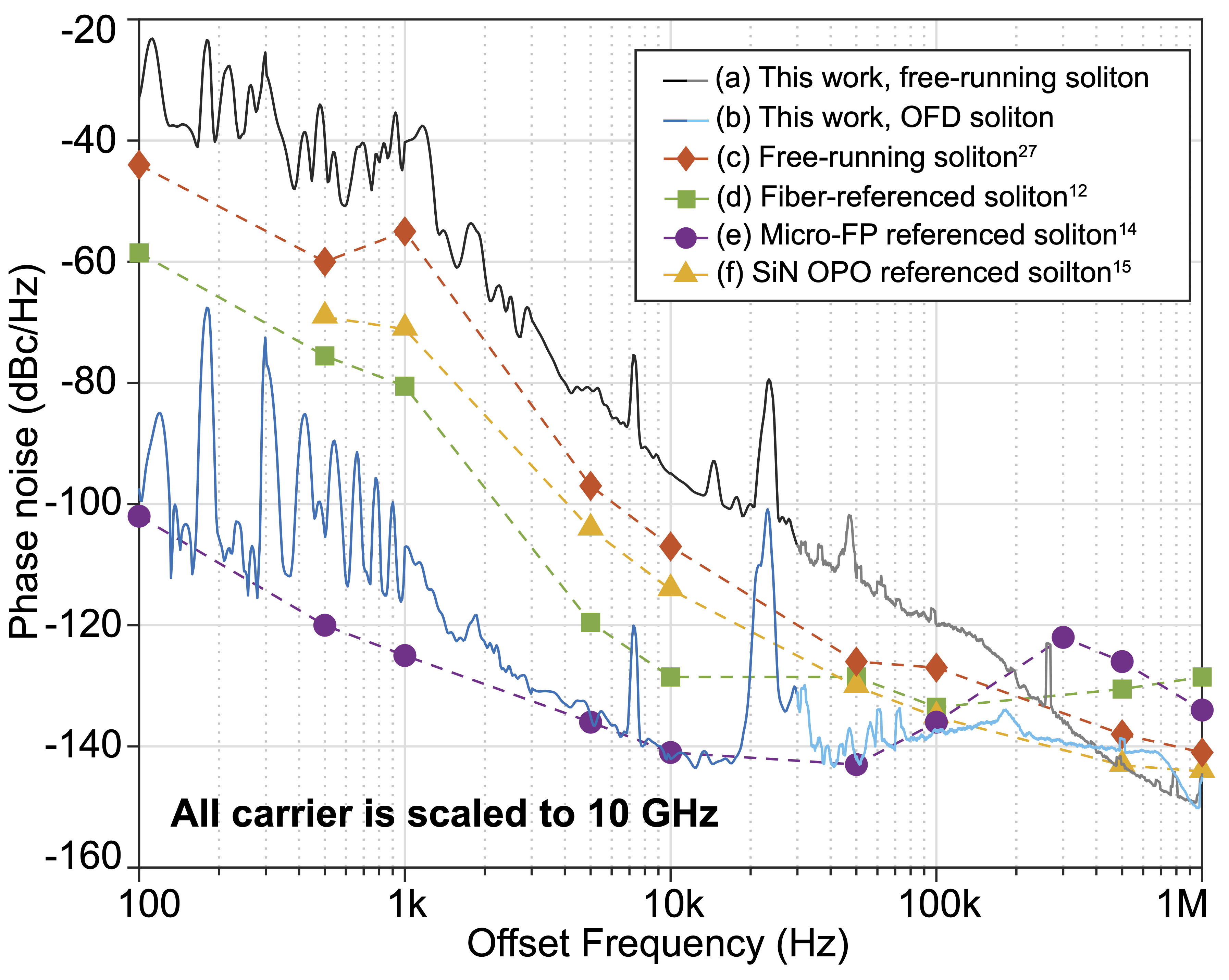}
\caption{{\bf Approximate phase noise of several integrated microcomb-based microwave and mmWave oscillators.} For comparison, carrier frequency is scaled to 10 GHz for all traces. (a) Free-running soliton phase noise of this work. (b) Phase noise of Kerr-OFD soliton in this work. Original carrier frequency: 109.5 GHz. (c) The best free-running integrated soliton microcomb \cite{liu2020photonic}. Original carrier frequency: 10 and 20 GHz. (d) SiN soliton microcomb referenced to 75-meter-long optical fiber \cite{tetsumoto2021optically}. Original carrier frequency: 300 GHz. (e) trace: integrated OFD oscillator referenced to microfabricated FP cavity \cite{kudelin2023photonic}. Original carrier frequency: 20 GHz. (f) soliton microcomb referenced to integrated OPO via all optical frequency division\cite{zhao2023all}. Original carrier frequency: 227 GHz.}
\label{fig:5}
\end{figure}

The soliton microcomb is generated in an integrated $Si_3N_4$ racetrack-shape microresonator, with a cross-section of 2.5 $\mu$m width $\times$ 0.8 $\mu$m height. The resonator has a free spectral range (FSR) of 109.5 GHz and an intrinsic (loaded) quality factor of 5.5 $\times 10^{6}$ (3.7 $\times 10^{6}$) at 1551 nm. A fast modulation sideband of laser A (1551 nm) serves as the pump of the soliton microcomb\cite{stone2018thermal}. The pump frequency can be rapidly tuned by tuning the voltage-controlled oscillator (VCO) that drives the modulator. The Laser B at 1600 nm with sub-mW power level is combined with the pump laser on a wavelength division multiplexer (WDM) and is injected into the microresonator. The optical spectrum of soliton and the injection laser B is shown in Fig.\ref{fig:Kerr_Lock}a. 

The frequency of injection laser B can be tuned continuously around 1600 nm to investigate the Kerr OFD mechanism (Fig.\ref{fig:Kerr_Lock}b). Fig.\ref{fig:Kerr_Lock}c shows the photodetector signal of the transmitted injection laser B and the soliton microcomb versus the frequency of laser B. Around -1.86 GHz offset frequency, the frequency difference between laser B and comb line $N=-54$ is within the photodiode bandwidth, and the beat note between laser B and comb line can be observed. When the $N$-th comb line is locked to the injection laser, the beat note oscillator disappears and is reduced to a DC voltage (area marked by orange). The instantaneous frequency of the beat note in the locking area is measured and shown in Fig.\ref{fig:Kerr_Lock}d. Between frequency offset -44 MHz to -109 MHz in Fig.\ref{fig:Kerr_Lock}d, no beat note frequency can be found, which indicates that comb line $N=-54$ is locked to the injection laser within this frequency range. Out of this range, the beat note frequency and its harmonics can be seen. The cavity resonance mode can also be observed in the transmission in Fig.\ref{fig:Kerr_Lock}c, which is near -0.57 GHz offset frequency. This unveils that strong Kerr locking can happen when the injection laser frequency is far from the cavity resonance frequency.

The Kerr locking leads to an optical frequency division, as $f_N=f_p + N \times f_r = f_{B}$, and $f_r = (f_{B}-f_p)/N$, where $f_{B}$ is the injection laser frequency (laser B), $f_{p}$ is the pump frequency, $f_N$ is $N$-th comb line frequency, $N$ is comb line number with respect to pump, and $f_{r}$ is rep-rate. The Kerr OFD can be observed by measuring the repetition rate versus the injection laser frequency (Fig.\ref{fig:Kerr_Lock}e). When the injection laser frequency is in the range from -44 MHz to -109 MHz, the soliton repetition rate changes linearly with the injection laser frequency. A line with a slope of $-1/54$ is plotted to show the division ratio. Finally, the Kerr OFD is measured under a series of the injected optical power of laser B, and the single-sided locking range versus injected power is plotted in Fig.\ref{fig:Kerr_Lock}f. The Kerr locking range can also be calculated analytically (see Methods), and we have: 
\begin{equation}
    |\omega_{B} - \omega_{s,N}| \leq 2 D_2 N^2 \frac{\kappa_e}{\kappa} \frac{\sqrt{P_{s,N} P_{B}}}{P_{s}}\stackrel{\text{def}}{=} \delta_{max},
    \label{eq:locking-range}
\end{equation}
where $\omega_{B}$ is the injection laser frequency, $\omega_{s,N}$ is the frequency of the $N$-th soliton comb line, $D_2$ is the microresonator dispersion, $\kappa$ is the resonator dissipative rate, $\kappa_e$ is the coupling rate between the microresonator and the waveguide. $P_{B}$ is the on-chip injection laser power, $P_s$ and $P_{s,N}$ are the soliton microcomb power and its $N$-th comb line power in the output waveguide. The measurement results show that the locking range increases with the square root of the injection laser power, which agrees with eq. (\ref{eq:locking-range}). Furthermore, the locking bandwidth of Kerr OFD can be calculated analytically under the small signal approximation:
\begin{equation}
    \delta_{BW} = \delta_{max} |\cos{\bar{\Theta}}|,
\end{equation}
where $\sin{\bar{\Theta}} = (\omega_{B} - \omega_{N,s})/\delta_{max}$ 
The maximum locking bandwidth equals the one-sided locking range. The locking bandwidth is important to the phase noise performance of Kerr OFD. The calculation also shows that the Kerr OFD has a 20 dB per decade gain (see eq.\ref{eq:phasenoise} in Methods), which is the same as the type-I locking loop.

The Kerr OFD can be used to generate ultra-stable microwave and mmWaves when the pump laser and the injection laser are stabilized to integrated optical coil references. 
Here, the reference cavity is an integrated thin-film SiN cavity with a cross-section of 6 $\mu$m width $\times$ 80 nm height. The cavity is 4-m long and coiled on a chip of cm-scale chip \cite{liu202236} with measured loaded quality factors of $46\times 10^{6}$ and $47\times 10^{6}$ at 1551 nm and 1600 nm, respectively. The large mode volume and high-quality factor together provide outstanding phase stability\cite{liu202236}. Both the pump laser and the injection laser are stabilized to the same reference cavity by using the PDH locking technique. The reference cavity is packaged to isolate technical noise from the environment. It is worth pointing out that the FSR of our coil reference cavity is $50$ MHz, which is smaller than the two-sided Kerr OFD locking bandwidth (65 MHz). As a result, there is always an optical mode of the reference cavity within the locking range to stabilize the injection laser.

To generate stable mmWave at 109.5 GHz, the Kerr OFD solitons are then optically amplified and illuminated on a high-speed charge-compensated modified uni-traveling carrier photodiode (CC-MUTC PD)\cite{xie2014improved,wang2021towards}. Since the carrier frequency is well above the frequency range of our phase noise analyzer, a mmWave to microwave frequency division (mmFD) method is applied to further divide the generated 109.5 GHz mmWave down to 18.1 GHz. 
Details of the mmFD method have been described elsewhere\cite{sun2023integrated}. After the mmFD, the spectra and phase noise of the 18.1 GHz signal can be measured directly on the phase noise analyzer. The electrical spectra from both the free-running soliton (orange trace) and the Kerr-OFD soliton (blue trace) are shown in the top-right inset of Fig.\ref{fig:phase_noise}b. 
The phase noise of the 18.1 GHz signal is shown in green in Fig.\ref{fig:phase_noise}b. Scaling the carrier frequency up to 109.5 GHz provides an upper bound limit for the phase noise of the 109.5 GHz mmWave, and both the free-running soliton (orange) and Kerr OFD soliton (blue) results are shown in Fig.\ref{fig:phase_noise}b. 
For general comparison with other oscillators, the Kerr OFD result is also scaled down to a typical 10 GHz carrier frequency and plotted in the dashed gray line. It should be noted that since the mmFD has a servo bandwidth of 312 kHz, phase noise measurement at high offset frequencies is limited by the mmFD method itself instead of the Kerr OFD. To evaluate the soliton phase noise at higher offset frequencies, a dual-tone delayed self-heterodyne interferometry method\cite{kwon2017reference} is used separately, and the combined phase noise is shown in Fig. \ref{fig:5}. 

The phase noise of the generated microwave and mmWave is exceptionally low. When the carrier frequency is scaled down to 10 GHz, the phase noise reaches -142 dBc/Hz at 10 kHz offset, which is a record-low value for any integrated photonic microwave and mmWave oscillators. Our reported value is approximately 27 dB better than the Keysight PSG standard model commercial signal generator. At 100 Hz offset frequency, the phase noise reaches an impressive -98 dBc/Hz. 
In figure \ref{fig:5}, we compare our results with the state-of-the-art integrated photonic oscillators. The orange trace shows the best integrated free-running soliton microwave oscillator\cite{liu2020photonic}. Other oscillators shown in figure \ref{fig:5} are all OFD-based, and they are referenced to integrated OPO \cite{zhao2023all} (orange), 75-meter long optical fiber \cite{tetsumoto2021optically} (green), and microfabricated Fabry–Pérot cavity \cite{kudelin2023photonic} (purple), respectively. Because of the high stability and low thermorefractive noise limit (TRN) of the reference cavity and the power of Kerr optical frequency division, our oscillator outperforms the free-running solitons and the integrated OPO-referenced solitons by several orders of magnitude and also outperforms the fiber-referenced OFD soliton. 
When compared to the best integrated photonic microwave oscillator reported to date \cite{kudelin2023photonic}, our oscillator performs equally well from 5 kHz to 100 kHz offset frequency. Above 100 kHz offset frequency, our oscillator shows lower phase noise because of the large locking bandwidth of the Kerr OFD. In comparison, there is a servo bump at 300 kHz offset frequency for the FP-referenced oscillator. Below 5 kHz, our phase noise is higher, and we suspect that this is because our soliton microcomb setup and the MUTC-PD setup are in two different rooms due to lab space limitation, and around 60 meter long fiber is used to connect them. This could add technical noise to the low offset frequency.

Finally, The mmWave power is characterized and shown in Fig.\ref{fig:phase_noise}d. Maximum power of 3.4 dBm is obtained at a photocurrent of 14 mA. The mmWave phase noise at different photocurrents is also recorded. Results at 1 kHz offset (light blue squares) and 10 kHz offset (dark blue dots) are plotted in Fig.\ref{fig:phase_noise}e, and the phase noise remains stable at different power levels. 
For comparison, the phase noise of Keysight PSG signal generator (standard model) is shown in the same figure in dashed lines, and the carrier frequency has been scaled up to 109.5 GHz.

In summary, we have demonstrated record-low phase noise mmWave generation with integrated photonics using integrated coil-reference cavity stabilized lasers and the Kerr OFD approach. Compared with conventional servo-OFD, the Kerr OFD is much simpler to implement and has a much larger locking bandwidth of tens of MHz. In this work, the phase noise of the generated mmWave at high offset frequency is limited by the reference laser, since the PDH locking loop in the reference laser has a bandwidth of only 145 kHz. This performance can be dramatically improved in the future by using optical reference based on stimulated Brillouin laser (SBS) \cite{lee2012chemically,gundavarapu2019sub,liu2024integrated}. Increasing the division ratio $N$ can further improve the phase noise performance, and this can be achieved by moving the injection laser frequency further away from the soliton center. While the lower comb line power at larger $N$ may decrease the Kerr OFD locking bandwidth \cite{wildi2023sideband}, this can be overcome by introducing dispersive waves to boost the comb line power by several orders of magnitude\cite{brasch2016photonic,yang2016spatial}. Finally, only 185 $\mu$W of the injection laser power is used in our low-phase noise Kerr OFD oscillator and no amplification of the injection laser is necessary. This shows that optical amplifier can be eliminated in low noise integrated OFD oscillators in the future when the Kerr OFD is combined with soliton microcombs that are directly pumped by laser diodes\cite{stern2018battery,xiang2021laser}.

\medskip

\noindent\textbf{Methods}
\begin{footnotesize}

\noindent{\bf Theoretical calculations of Kerr OFD.} 
The Kerr OFD system consists of soliton microcombs and an additional laser injected into the soliton-forming Kerr micoresonator. This can be described by the modified LLE equation \cite{yang2017counter,wildi2023sideband}, and the locking range has been solved analytically using the Lagrangian perturbation method\cite{wildi2023sideband}. The modified LLE can be solved analytically using the alternative momentum method \cite{yang2017counter}.

The equation of motion of the soliton field can be expressed as: 

\begin{equation}
\begin{split}
    \frac{dA(\phi, t)}{dt} = i\frac{D_2}{2} \frac{\partial^2 A}{\partial \phi^2} + ig|A|^2 A - i\delta \omega A - \frac{\kappa}{2} A + \sqrt{\kappa_e P_{p}}\\ + ig\tau_R D_1 A \frac{\partial |A|^2}{\partial \phi} + \sqrt{\kappa_e } b e^{iN\phi},
    \end{split}
    \label{eq:LLE}
\end{equation}
where $A(\phi,t)$ is the optical field of the soliton in the microcavity, $\phi$ is the angular coordinate in the rotational frame, $D_2$ is the coefficient associated with group velocity dispersion, $g$ is the normalized Kerr nonlinear coefficient, $\delta \omega$ is the cavity-laser detuning, $\kappa$ is the dissipative rate of the resonator, $\kappa_e$ is the coupling rate between the resonator and the waveguide, $P_p$ is the power of the pump laser, $\tau_R$ is the time constant of Raman effect, $D_1$ is the free-spectral-range (FSR) at the pumping mode, $b$ is the field of the laser injected into the soliton-forming microresonator, which is roughly $N$ FSR away from the pump laser. $|A|^2$ and $|b|^2$ are normalized to optical energy and optical power, respectively. By following the same procedure as in counter-propagating soliton locking calculation\cite{yang2017counter}, the following equation can be derived from equation (\ref{eq:LLE}), 
\begin{equation}
\begin{split}
    \frac{\partial \mu_c}{\partial t} &=  - \kappa \mu_c + \kappa \mu_R + \frac{N\sqrt{\kappa_e}}{E_s} (a_N b^* + a_N^* b)\\ &= - \kappa \mu_c + \kappa \mu_R + \frac{2N\sqrt{\kappa_e}}{E_s} |a_N b| \sin{\Theta},
    \end{split}
\end{equation}
where $\mu_c$ is the mode number of the soliton spectral center, $\mu_R$ is the soliton spectral center shift induced by the Raman self-frequency shift effect\cite{yi2016theory}, and $\mu = 0$ corresponds to the mode being pumped by the pump laser. $E_s$ is the intracavity energy of soliton. $a_N$ is the intracavity field of the $N$-th comb line of the soliton, and $\Theta = \theta_{s,N} - \theta_{B} + \pi/2$ is the phase difference between the $N$-th soliton comb line ($a_N$) and the injection optical field ($b$). Note that $b = \sqrt{P_B} e^{-i\Delta \omega t}$, where $\Delta \omega = \omega_{B} - (\omega_p + D_1 N)$ is the injection laser frequency in the rotation frame. Similarly, $a_N = |a_N| e^{-i(\omega_{N,s} - \omega_p - D_1N)t + i\varphi}$, where $\omega_{N,s}$ is the frequency of the $N$-th soliton comb line, $\varphi$ is the overall phase of the soliton envelope, and $\omega_{N,s} = \omega_p + N\omega_r = \omega_p + N (D_1 + D_2 \mu_c)$. Therefore, $\frac{\partial \Theta}{\partial t}$ can be expressed as $\frac{\partial \Theta}{\partial t} = \Delta \omega - N D_2 \mu_c$. Take the derivative of this relationship, and we can arrive at

\begin{equation}
\begin{split}
    \frac{\partial^2 \Theta}{\partial t^2} + \kappa \frac{\partial \Theta}{\partial t}  &= \kappa \delta - \frac{2N^2 D_2 \sqrt{\kappa_e}}{E_s} |a_N b| \sin{\Theta} \\ &= \kappa \delta - \frac{2\kappa_e N^2 D_2}{P_{s}} \sqrt{P_{s,N}P_B} \sin{\Theta},
    \end{split}
    \label{eq:theta}
\end{equation}
where $\delta = \omega_{B} - \omega_{N,S}$ is the frequency difference between the injection laser and the $N$-th line of the free-running soliton. Using input-output relationship, the intracavity field ($a_N$) and energy ($E_s$) can be converted to optical power in the output waveguide, which can be measured directly. $P_s$ and $P_{s,N}$ are the optical power of the soliton microcombs and its $N$-th comb line in the output waveguide, respectively.

\medskip

\noindent {\bf Kerr OFD locking range:} The locking range of Kerr OFD can be obtained from the steady-state solution of equation \ref{eq:theta}. At steady-state, equation \ref{eq:theta} has solution of: 
\begin{equation}
     \delta = \frac{2\kappa_e N^2 D_2}{\kappa P_{s}} \sqrt{P_{s,N}P_B} \sin{\Theta} 
\end{equation}
Since $|\sin{\Theta}| \leq 1$, the steady-state solution only exists when
\begin{equation}
    |\delta| \leq 2 N^2 D_2\frac{\kappa_e}{\kappa} \frac{\sqrt{P_{s,N}P_B}}{P_{s}}\stackrel{\text{def}}{=} \delta_{max}.
\end{equation}
The locking range of Kerr OFD can then be found as $|\omega_{B} - \omega_{N,S}| = |\delta| \leq \delta_{max}$.

\noindent {\bf Kerr OFD bandwidth and gain:} The locking bandwidth and gain are important metric for optical frequency division. Both of them can be analytically calculated by applying small signal analysis to equation (\ref{eq:theta}). In the small signal approximation, we can define $\delta = \bar{\delta} + \Delta \delta$ and $\Theta = \bar{\Theta} + \Delta \theta$, where $\bar{\delta}$ and $\bar{\Theta}$ are the mean value of $\delta$ and $\Theta$, respectively. Equation (\ref{eq:theta}) then becomes
\begin{equation}
    \frac{\partial^2 \Delta \theta}{\partial t^2} + \kappa \frac{\partial \Delta \theta}{\partial t}  = \kappa \Delta \delta - \kappa \delta_{max} \Delta \theta \cos{\bar{\Theta}}.
    \label{eq:theta-small}
\end{equation}
From $\Theta = \theta_{s,N} - \theta_{B} + \pi/2$, we can have $\Delta \theta = \Delta \varphi_p + N \times \Delta \varphi_r - \Delta \varphi_B$, where $\Delta \varphi_p$, $\Delta \varphi_{B}$ and $\Delta \varphi_r$ are the small phase variation of the pump laser, the injection laser, and soliton repetition rate, respectively. Similarly, $\Delta \delta = \Delta \omega_{B} - \Delta \omega_p - N \Delta \omega_r^{free}$, where $\Delta \omega_p$, $\Delta \omega_{B}$, and $\Delta \omega_r^{free}$ are the small frequency variation of the pump laser, the injection laser, and the free-running soliton repetition rate. To calculate the phase noise of the soliton repetition rate in Kerr OFD, we can apply Fourier transform to equation (\ref{eq:theta-small}) and use the general relationship between phase and frequency ($\omega = - \partial \varphi/\partial t$) to arrive at: 
\begin{equation}
\begin{split}
    \tilde{\varphi}_r(s) = (1+ \frac{i\kappa s}{s^2- i\kappa s - \kappa \delta_{max} \cos{\bar{\Theta}}}) \times \frac{\tilde{\varphi}_{B}(s) -\tilde{\varphi}_{p}(s)}{N} \\ + \frac{i \kappa s}{s^2- i\kappa s - \kappa \delta_{max} \cos{\bar{\Theta}}} \times \tilde{\varphi}_r^{free}(s),
\end{split}
\end{equation}
where $\tilde{\varphi}(s)$ is the Fourier transform of phase $\Delta \varphi$, and $s$ is offset frequency from the carrier. As soliton repetition rate phase noise is $L_r(s) = |\tilde{\varphi}_r(s)|^2$, we arrive at: 
\begin{equation}
\begin{split}
    L_r(s) = |1+ \frac{i\kappa s}{s^2- i\kappa s - \kappa \delta_{max} \cos{\bar{\Theta}}}|^2 \times \frac{L_{B}(s) -L_{p}(s)}{N^2} \\ + \frac{1}{1 + |\delta_{max} \cos{\bar{\Theta}}/s - s/\kappa|^2} \times L_r^{free}(s). 
\end{split}
\label{eq:phasenoise}
\end{equation}
The first term on the right of the equation gives the $1/N^2$ phase noise reduction factor in optical frequency division. The second term on the right of the equation gives the gain that suppresses the phase noise of the free-running soliton repetition rate. In the case of $s \ll \kappa$, the Kerr OFD gain reduces 20 dB per decade with the increase of the offset frequency $s$, until the offset frequency approaches $\delta_{max} \cos{\bar{\Theta}}$. $\kappa/2\pi \approx 52$ MHz in this work. Therefore, the Kerr OFD servo bandwidth can be approximate to $\delta_{BW} = \delta_{max} |\cos{\bar{\Theta}}|$. Another observation is that the 20 dB/decade change in gain is identical to the type I phase lock loop.


\end{footnotesize}

\medskip

\noindent\textbf{Acknowledgement}

\noindent The authors acknowledge Madison Woodson and Steven Estrella from Freedom Photonics for MUTC-PD fabrication, Ligentec for SiN microresonator fabrication, and gratefully acknowledge DARPA GRYPHON (HR0011-22-2-0008), National Science Foundation (2023775), Advanced Research Projects Agency—Energy (DE-AR0001042), Air Force Office of Scientific Research (FA9550-21-1-0301). The views and conclusions contained in this document are those of the authors and should not be interpreted as representing official policies of DARPA, ARPA-E, or the U.S. Government.

\medskip

\noindent \textbf{Author Contributions}\\ 
X.Y., S.S. designed the experiments. S.S., F.T., and S.H. performed the system measurements. M.W.H., K.L., J.W., D.J.B., K.D.N. and P.A.M. developed the reference lasers. J.S.M. and A.B. designed and fabricated the CC-MUTC PDs. S.S., X.Y., F.T., and S.H. analyzed the experimental results. All authors participated in preparing the manuscript.

\medskip

{\noindent \bf Competing interests}
The authors declare no competing interests.

\medskip

{\noindent \bf Data availability.} The data that support the plots within this paper and other findings of this study will be made available online before publication. 

\medskip

{\noindent \bf Code availability.} The codes that support the findings of this study are available from the corresponding authors upon reasonable request.

\bibliographystyle{naturemag}
\bibliography{ref}

\end{document}